\documentstyle[sprocl]{article}

\input psfig.sty

\def\be{\begin{equation}}
\def\ee{\end{equation}}
\def\bea{\begin{eqnarray}}
\def\eea{\end{eqnarray}}

\def\c12ag{$^{12}C(\alpha,\gamma)^{16}O$}
\def\xn13pg{$^{13}N(p,\gamma)^{14}O$}
\def\be7pg{$^7Be(p,\gamma)^8B$}
\def\o14{$^{14}O$}
\def\xbe7{$^7Be$}
\def\b8{$^8B$}
\def\n16{$^{16}N$}

\begin{document}

\hspace{3in} {UConn-40870-0020}

\title{PROGRESS IN NUCLEAR ASTROPHYSICS USING SECONDARY-RADIOACTIVE BEAMS
\footnote{Work Supported by USDOE Grant No. DE-FG02-94ER40870.}}

\author{Moshe Gai
\footnote{Invited Talk, Int. Conf. in memory of C. S. Wu,
 Aug. 16-18, 1997, Nanjing, China}}

\address{Dept. of Physics, U46, University of Connecticut,
   2152 Hillside Rd., \\ Storrs, CT 06269-3046, USA; 
   gai@uconnvm.uconn.edu, http://www.phys.uconn.edu}

%%%%%%%%%%%%%%%%%%%%%%%%%%%%%%%%%%%%%%%%%%%%%%%%%%%%%%%%%%%%%% 
% You may repeat \author \address as often as necessary	%
%%%%%%%%%%%%%%%%%%%%%%%%%%%%%%%%%%%%%%%%%%%%%%%%%%%%%%%%%%%%%% 

\maketitle

\abstracts{We review progress in studying two central 
problems in Nuclear Astrophysics: the \be7pg reaction rate
at very low energies, of importance for  
estimating the Solar Neutrino flux, and the \c12ag 
reaction rate, of importance for stellar processes in 
a progenitor star prior to a super-nova collapse. \\
The \be7pg reaction is one of the major source of uncertainties 
in estimating the \b8 solar neutrino flux and is critical for  
the Solar Neutrino Problem. The main source of uncertainty is the existence
of conflicting data with different absolute normalization.
While attempts to measure this reaction rate with \xbe7 beams are 
under way we discuss a newly emerging method to extract this cross 
section from the Coulomb dissociation of the radioactive beam of \b8.
We discuss some of the issues relevant for this study including the 
question of the E2 contribution to the Coulomb dissociation process 
which was recently measured to be small.
The Coulomb dissociation appears to provide a viable alternative method 
for measuring the \be7pg reaction rate.\\
Several attempts to constrain the 
p-wave S-factor of the \c12ag reaction at
Helium burning temperatures (200 MK) using the beta-delayed alpha-particle 
emission of \n16 have been made, and it is claimed that this 
S-factor is known, as quoted by the TRIUMF collaboration. 
In contrast reanalyses (by G.M. hale) of all thus far available data (including 
the \n16 data) does not rule out a small S-factor solution. 
Furthermore, we improved our previous Yale-UConn 
study of the beta-delayed alpha-particle 
emission of \n16 by improving our statistical sample
(by more than a factor of 5), improving 
the energy resolution of the experiment (by 20\%), and in understanding
our line shape, deduced from measured quantities.  Our newly measured 
spectrum of the beta-delayed alpha-particle emission of \n16
is not consistent with the TRIUMF('94) data, but  
is consistent with the Seattle('95) data, as well as the 
earlier (unaltered !) data of Mainz('71). The implication 
of this discrepancies for 
the extracted astrophysical p-wave s-factor is briefly discussed.}

\section{The Coulomb Dissociation of $^8B$ and the \be7pg Reaction
     at Low Energies}

The Coulomb Dissociation \cite{Bau86} is a
Primakoff \cite{Pr51} process that could be viewed in first order as the time
reverse of the radiative capture reaction.  In this case instead of studying
for example the fusion of a proton plus a nucleus (A-1), one studies the
disintegration of the final nucleus (A) in the Coulomb field, to a proton
plus the (A-1) nucleus.  The reaction is made possible by the absorption of
a virtual photon from the field of a high Z nucleus such as $^{208}Pb$.  In this
case since $\pi/k^2$  for a photon is approximately 1000 times larger than that
of a particle beam, the small cross section is enhanced.  The large virtual
photon flux (typically 100-1000 photons per collision) also gives rise to
enhancement of the cross section.  Our understanding of the Coulomb
dissociation process \cite{Bau86} allow us to extract the inverse
nuclear process even when it is very small.  However in Coulomb
dissociation since $\alpha Z$  approaches unity (unlike the case in electron
scattering), higher order Coulomb effects (Coulomb post acceleration) may
be non-negligible and they need to be understood 
\cite{Ber94,Typ94}.  The success of
the experiment is in fact contingent on understanding such effects and
designing the kinematical conditions so as to minimize such effects.

Hence the Coulomb dissociation process has to be measured with great care
with kinematical conditions carefully adjusted so as to minimize nuclear
interactions (i.e. distance of closest approach considerably larger then 20
fm, or very small forward angles scattering), and measurements must be
carried out at high enough energies (many tens of MeV/u) so as to maximize
the virtual photon flux.

The Coulomb dissociation of $^8B$ may provide a good opportunity for resolving
the issue of the absolute value of the cross section of the $^7Be(p,\gamma)^8B$
reaction.  The Coulomb dissociation yield
arise from the convolution of the inverse nuclear cross section times the
virtual photon flux.  While the first one is decreasing as one approaches
low energies, the second one is increasing (due to the small threshold of
137 keV).  Hence over the energy region of 400 to
800 keV the predicted measured yield is roughly constant.  This is in 
contrast to the case of the nuclear cross section that is dropping very
fast at low energies.  Hence measurements at these energies
could be used to evaluate the absolute value of the cross section.

An experiment to study the Coulomb dissociation of $^8B$  was performed 
at the {\bf RIKEN-RIPS radioactive beam facility} \cite{Mo94}.
Indeed the results of the experiment allow us to measure the cross section of 
the $^7Be(p,\gamma)^8B$ radiative capture reaction and preliminary results 
are consistent with the absolute value of the cross section 
measured by Filippone et al. \cite{Fi83} and by Vaughn et al. \cite{Va70}, 
but not Kavanagh \cite{Ka69} and Parker \cite{Pa66}, 
as shown in Fig. 1.

\centerline{\psfig{figure=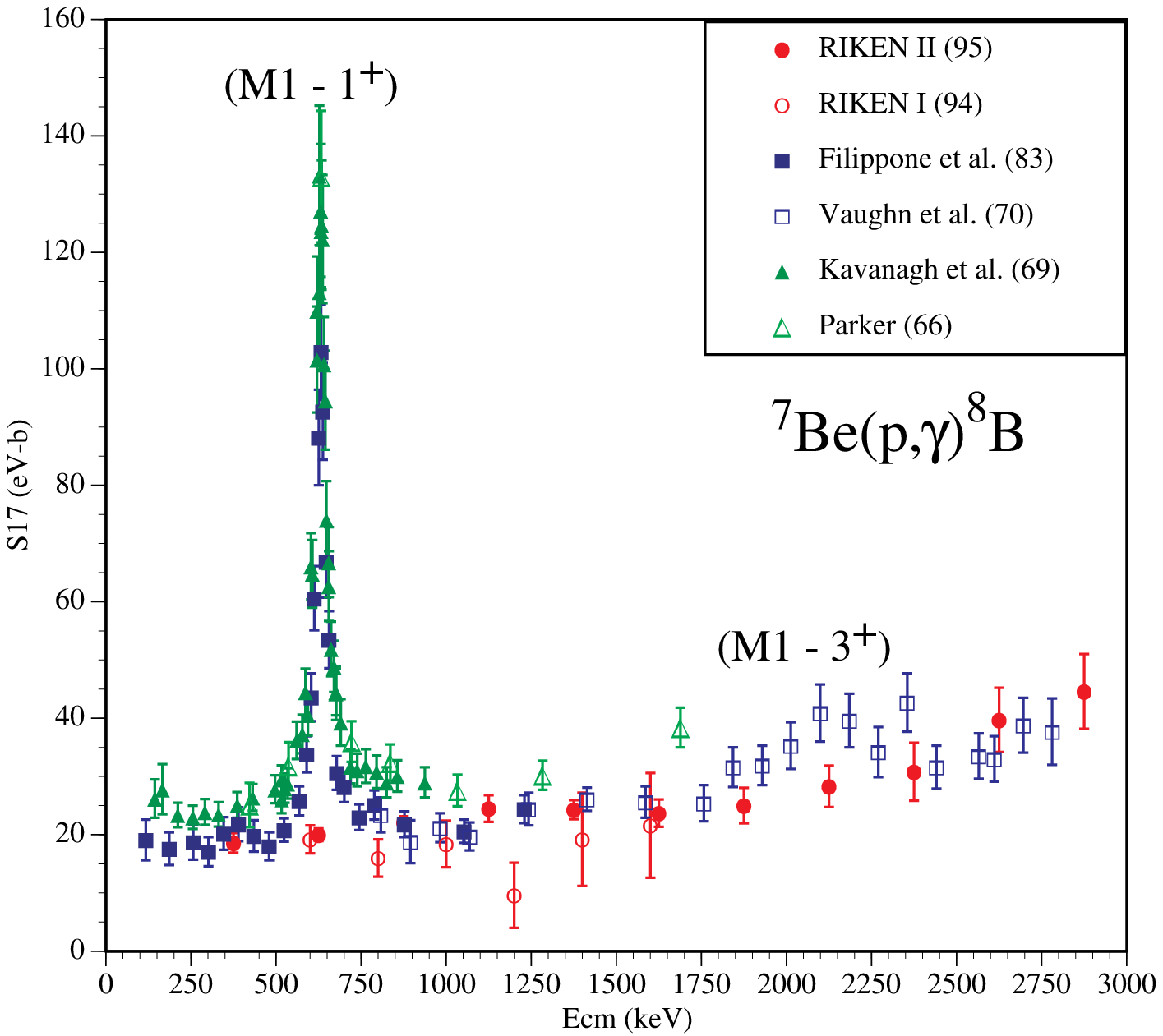,height=3.0in}}

\begin{center}
\underline{Fig. 1:} Measured values of $S_{17}(E)$.
\end{center}

\subsection{Is There Evidence for an E2 Component?}

A search for E2 component in the RIKEN data \cite{Mo94} 
was performed by Gai
and Bertulani \cite{Ga94}. When the experimental
resolutions are correctly taken into account, together with the
correct RIKEN data, the best fit of the angular distributions is obtained
with E1 amplitude alone. Our analysis invalidates
previous claims \cite{Lang}.

In addition we have measured in a separate experiment 
\cite{Ki97} detailed angular 
distributions for the Coulomb dissociation of \b8 in an attempt to 
extract the E2 amplitude directly.
The $^{208}$Pb target and $^8$B beam properties 
in this experiment were as in Ref. \cite{Mo94},
but the detector system covered a large angular range up to around 9$^{\circ}$
to be sensitive to the E2 amplitude.
The E1 and E2 virtual photon fluxes were calculated 
\cite{Ki97} using quantum mechanical approach.
The nuclear amplitude is evaluated based on the collective form factor
where the deformation length is taken to be the same as the Coulomb one.
This nuclear contribution results in possible uncertainties in the fitted
E2 amplitude. Nevertheless, the present results lead to a very small
E2 component at low energies, below 1.5 MeV, of the order of a few
percent, even smaller than the low value predicted by Typel and 
Baur \cite{Typ94}.  A recent reanalysis of the RIKEN2 data \cite{Ki97} 
by Bertulni and Gai \cite{Ber97} confirmed the 
small E2 extracted by Kikuchi et al. \cite{Ki97} as well as 
the negligible nuclear contribution.
Recently a possible mechanism to reduce the E2 dissociation amplitude
was proposed by Esbensen and Bertsch \cite{Esben}.

\subsection{Conclusions}

In conclusion we demonstrate that the Coulomb dissociation provides
a viable alternative method for measuring small cross section of
interest for nuclear-astrophysics. First results on the CD of \b8 are
encouraging for a continued effort to extract $S_{17}(0)$, of importance
for the SSM. Our initial results are consistent with
the lower value of the cross section measured by Filippone et al. and
suggest a small value for the extracted $S_{17}(0)$; smaller than
20 eV-barn, and considerably smaller (30\%) than assumed in the 
Standard Solar Model.

\section{Helium Burning: The \c12ag Reaction and the Beta-Delayed
      Alpha-Particle Emission of \n16}

In this section we discuss progress in studying the \c12ag reaction 
rate of importance for understanding 
helium burning \cite{Fo84} in massive stars.
We study this reaction in its time 
reverse process using the beta-delayed
alpha-particle emission of \n16, allowing us to add useful data 
and constraints on the reaction rate, and the
extraction of the p-wave astrophysical S-factor.
However, it appears that {\bf early hopes} for deducing the
p-wave astrophysical S-factor ($S_{E1}$) using the \n16
data {\bf are not substantiated}.  And further confusion is generated
by inconsistent data on the beta-delayed alpha-particle
emission of \n16 in addition to inconsistent data on the
\c12ag reaction.

We emphasize that while data on the beta decay of \n16 may
add useful constraint and may allow for extracting the (virtual) reduced 
alpha-particle width of the bound $1^-$ state, the sign of the
mixing phase of the bound and quasi-bound $1^-$ states in the
\c12ag reaction has nothing to do with the beta-decay of \n16 and
can not be directly determined from the data on the beta-delayed 
alpha-particle emission of \n16. It turns out that this difficulty
does not allow for unambiguous extraction of the p-wave S-factor
even with the inclusion of the new data on \n16.  Furthermore, 
a reanalysis of all existing data (including the \n16 data)
by Gerry Hale \cite{Ha96} demonstrates that a small S-factor 
solution could not be ruled out.  In fact Hale's best fit for the 
TRIUMF \n16 data \cite{Az94} is for an E1 S-factor approximately 
20 keV-b.  Interestingly, Hale's best fit is consistent with a broader 
line shape.  As we discuss below, such a broader line shape is observed 
in all other data sets, and it is quite possible that the 
narrow line shape of the TRIUMF data is an artifact of their data
analysis.

\subsection{The TRIUMF Result}

A measurement of the beta-delayed alpha-particle emission of \n16
was performed at TRIUMF \cite{Bu93,Az94}.  The spectrum is observed with
high statistics (approximately one million events) 
and indeed the TRIUMF collaboration claims to have deduced
the p-wave astrophysical S-factor with high accuracy.  
Based on for example, their R-matrix
analysis they quote a large value of:
$S_{E1}=81 \pm 21\ keV-b$. The E1 S-factor was 
previously uncertain by approximately a factor of 10 
and we note the relatively
high accuracy and the implication 
that they determined the interference of the two $1^-$ states in
\c12ag to be constructive (i.e. large S-factor).

As we demonstrate in this paper there is enough reasons to doubt the
TRIUMF data, and furthermore we do not confirm the conclusion of the
TRIUMF group that the p-wave S-factor of the \c12ag reaction has been
measured.

\subsection{The New Yale-UConn Experiment}

A further measurement of the Beta-Delayed Alpha-Particle energy 
spectrum of \n16 at low energy was performed in continuation of 
the first generation Yale experiment 
\cite{Zh93,Zh93a}.  The final phase of this experiment was performed 
using the Yale ESTU tandem van de Graaff accelerator at the Wright 
Laboratory at Yale University during the summer of 1995 \cite{Fr96,Fr96a}.

The \n16 was produced using a 70 MeV $^{15}N$ beam and a 1250 
Torr, 7.5 cm long deuterium gas target with 25 $\mu m$ beryllium 
entrance and exit foils.  The \n16 emerged from the gas target with a 
broad recoil energy spectrum, with the lower 1 MeV portion 
stopping in a thin (190 $\mu g/cm^2$) aluminum catcher foil tilted 
at 7$^\circ$ with respect to the beam.  After the \n16 was 
captured, the catcher foil was rotated 180$^\circ$ 
into the counting area.  While the arm rotated and the detectors 
counted, a tantalum beam chopper was used to block the beam far 
upstream.  Each full production and counting cycle lasted 21 
seconds, approximately twice the lifetime of \n16.

The counting area contained, as in our previous
experiment \cite{Zh93,Zh93a}, 9 thin 
Silicon Surface Barrier (SSB) detectors used to measure the energy 
and timing information of the alpha-particles in coincidence with 
an array of 12 fast plastic scintillator detectors, which measured 
the timing of beta-particles.  This timing information was used to 
reduce (by more than a factor of 100 over the low energy range 
of interest) the background in our SSB array due to detection of 
beta-particles and due to partial charge collection in the SSB 
detector.

The line shapes of both the first and second Yale-UConn data sets
are the same \cite{Fr96,Fr96a}.
In order to consider the line shape of both 
Yale-UConn data sets, it is useful to consider a situation for a 
predicted spectrum which is constant in energy (or time).  Clearly 
the yield at a specific energy (time) is directly proportional to 
the energy (time) resolution at that energy (time).  In this case 
the energy (time) resolution is the integration interval.  
Hence our data need to be divided by the varying energy 
resolution for alpha-particles traversing our aluminum foil 
and the time resolution of our time of flight system. 
 The time resolution of our experiment is 
measured directly in the data on the beta-delayed alpha-particle 
of \n16 as well as the beta-delayed alpha-particle emission of 
$^8Li$ which was also measured in our experiment using the same 
setup and the $^7Li(d,p)^8Li$ reaction.  Hence the line shape in 
the current (and previous) 
experiment(s) is deduced from measured $\partial E/\partial x$ data 
and the measured time resolution of our experiment.

\centerline{\psfig{figure=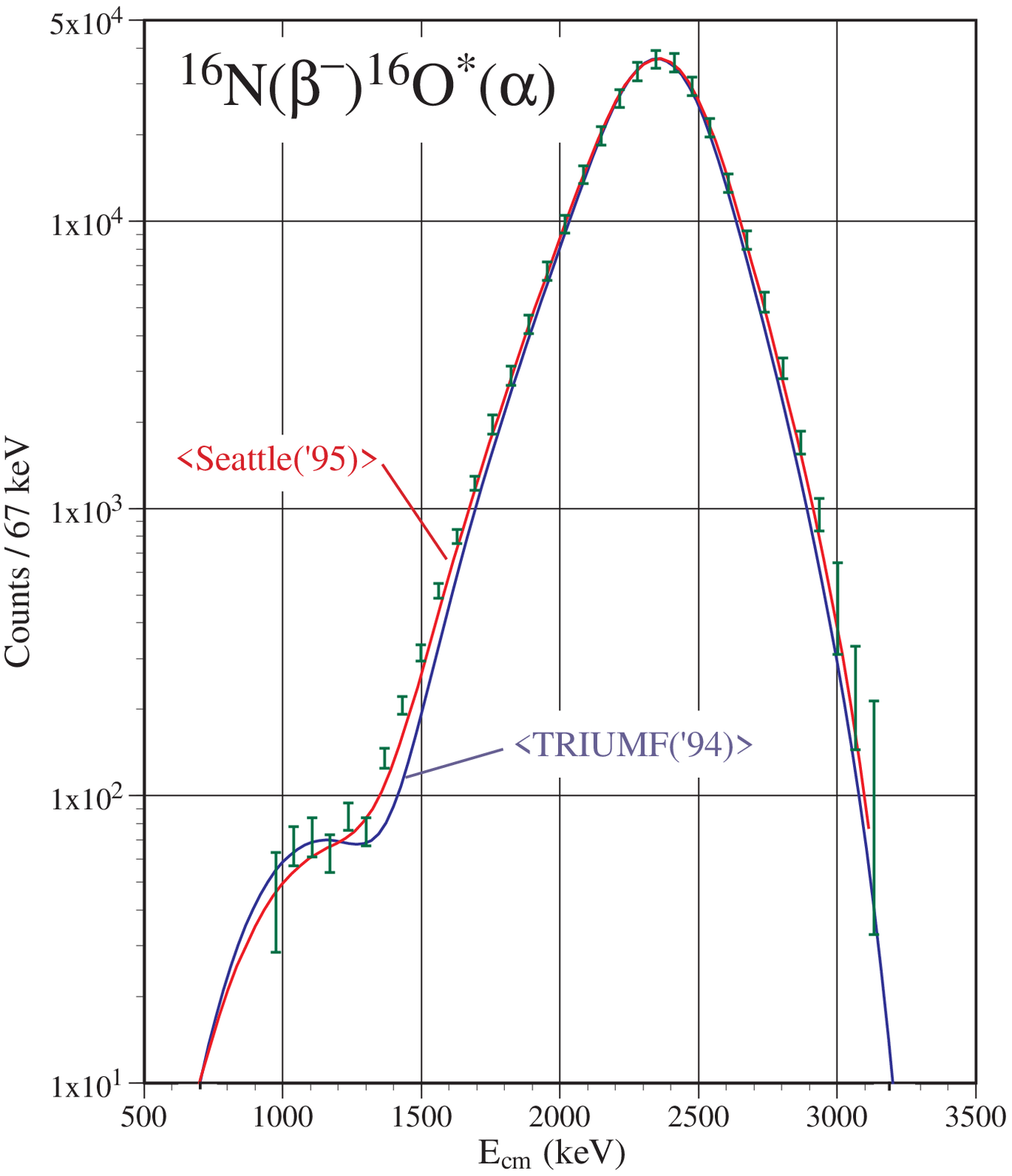,height=3.75in}}

\underline{Fig. 2:}  New Yale-UConn data, corrected for line shape,
compared with TRIUMF \cite{Az94} and Seattle \cite{Zh95} theory curves
(with reduced chi-squares).
The theory curves have been averaged over the 
experimental energy resolution.

We improved our previous Yale-UConn experiment \cite{Zh93,Zh93a} 
by: (1) A 20\% improvement of our energy resolution (200 keV at 2.36 MeV), 
(2) More than a factor of five increase in statistics (292,000 
events), and (3) An understanding of our line shape deduced from 
measured quantities.  Our results are shown in Fig. 2.  The data 
shown in Fig. 2 were corrected for the energy dependence of the 
$\beta -\alpha$ coincidence efficiency and line shape, both
deduced from measured quantities. The uncertainty of the three
highest energy points include the uncertainty of the
$\beta -\alpha$ coincidence efficiency.

\subsection{Comparison of TRIUMF data to other data sets}

In Fig. 2 we also show our data compared to the Seattle \cite{Zh95} and 
TRIUMF \cite{Az94}\ theory curves 
averaged over the variable energy resolution of our
experiment.  Note that the theory curves are a good representation 
of their respective data, but they allow us to carry out the 
energy averaging also over the edges of the finite data.  With the 
Seattle theory superimposed on our data we calculate a
$\chi ^2$\ per data point of 1.4 and for TRIUMF theory 7.2.  We conclude 
that our data confirm the Seattle data \cite{Zh95}\ but 
do not confirm the 
TRIUMF data \cite{Az94}.  Most notable is the absence 
of a well defined minimum at approximately 1.4 MeV as suggested by 
the TRIUMF data.  The data in the vicinity of 1.4 is dominated by 
the f-wave contribution and hence essentially determines the f-wave 
contribution.  A larger f-wave contribution (at 1.4 MeV) would 
naturally lead to a smaller p-wave contribution at the interference 
maximum (at 1.1 MeV) and thus a smaller p-wave astrophysical S-factor.

\centerline{\psfig{figure=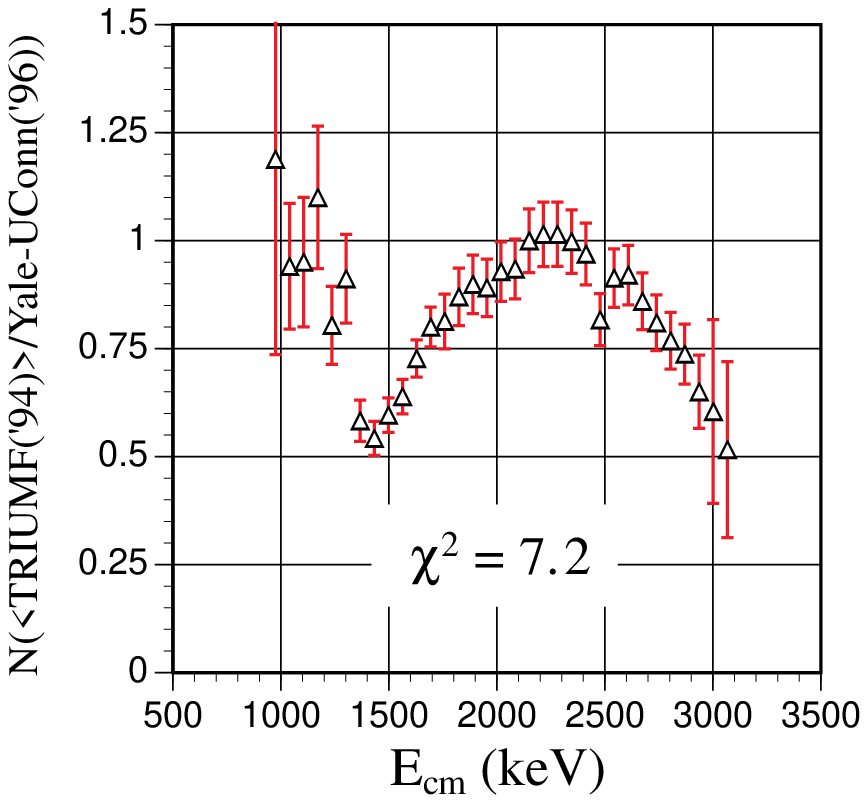,height=2.5in}}

\underline{Fig. 3:}  Ratio of the energy averaged TRIUMF theory 
curve  \cite{Az94} to the new Yale-UConn data.

\centerline{\psfig{figure=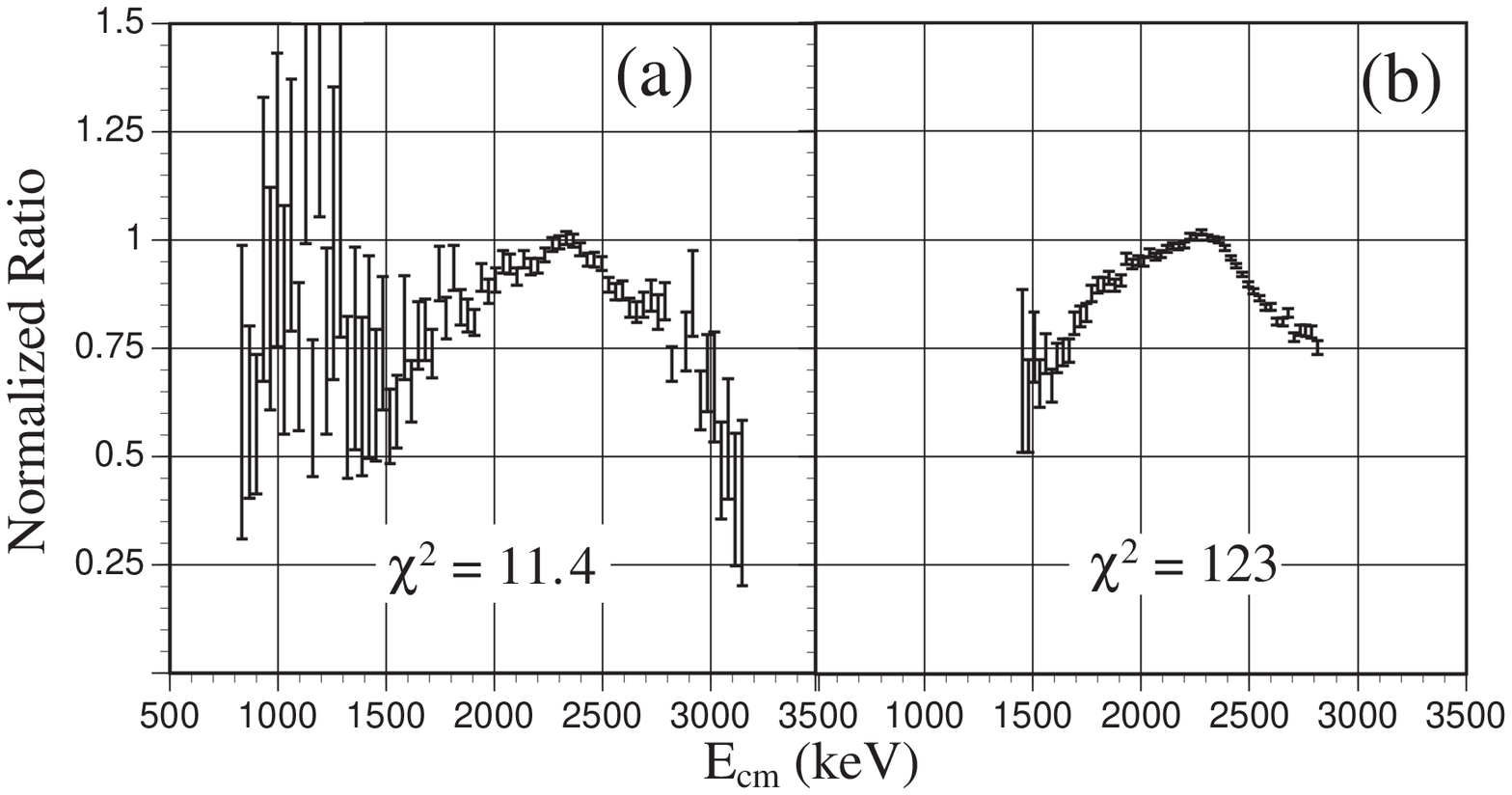,height=2.5in}}

\underline{Fig. 4:}  Ratios of (a) the TRIUMF data set \cite{Az94} to
the Seattle data set \cite{Zh95} and (b) the TRIUMF data set \cite{Az94} to
the Mainz data set \cite{Barker,Ne74}.  Linear interpolations were used
when necessary. Notice that the ratio plots are
very similar to each other and to the plot in Figure 3
indicating that there are
three data sets: the Mainz('71), 
Seattle('95), and the current Yale-UConn('96) that agree with each 
other but disagree with the TRIUMF('94) data.

Following the conclusion that our data is consistent with the
Seattle data but not TRIUMF data, as shown in Figs. 2 and 3, 
we received from Fred Barker 
\cite{Barker,Ne74} a copy of the original communication from Waffler to 
Barker dated 5 Feb. 1971, which includes approximately 32 million events 
and a measured beta-particle background spectrum. This data set was 
originally taken in a study of the parity violating alpha decay from the
8.8719 MeV $2 ^-$\ state in $^{16}O$.  We first note that we do not 
confirm \cite{Fra97} the allegation that there is a problem 
with the energy calibration of the Mainz data. We have in fact shown \cite{Fra97}
that the alteration of the Mainz data by the TRIUMF calibration can 
not be justified. Using the original unaltered Mainz data
we observe that it agree with the Seattle data ($\chi ^2$\ per
data point of 2.5) and disagrees with the TRIUMF data ($\chi ^2$\ per 
data point of 123).  In Fig. 4 we show using a
linear scale, the ratio of the
TRIUMF(94) data to other data sets.  
Note that the disagreement with the 
TRIUMF data in all cases is equally bad on the high and 
low energy sides of the main peak at 2.35 MeV.  This together with the 
fact that all data sets agree on the low energy interference
maximum, negates arguments of low energy tails.  
We conclude that indeed all other
data sets that were measured with the \n16 produced via the 
$^{15}N(d,p)^{16}N$ reaction 
including Mainz(71), Seattle(95) and Yale-UConn(96)
agree with each other and exhibit the (same) disagreement with the
TRIUMF(94) data.

\subsection{Comparison of TRIUMF(93) data to TRIUMF(94)}

This disagreement suggest two possible conclusions.  One, that all data other
than the TRIUMF data are wrong and only the TRIUMF data exhibit the true
narrow line shape. Second, that the narrow line shape of the TRIUMF(94) data
is an artifact of the coincidence data analysis.

In order to further investigate these two possibilities we have examined
the TRIUMF(93) data \cite{Bu93} as compared to TRIUMF(94) data \cite{Az94}
-- as reanalyze by the graduate student James Powell.
And in Fig. 5 we show the ratio of the TRIUMF(94) data to TRIUMF(93) data.
Clearly the TRIUMF(93) data exhibit yet even a narrower line shape than
TRIUMF(94). But the TRIUMF(93) data was already rejected by the TRIUMF
collaboration, as discussed in \cite{Az94}, and clearly this demonstrate
that the narrow line shape of the TRIUMF(93) data is an artifact of the
analysis (i.e. energy miscalibration).

\centerline{\psfig{figure=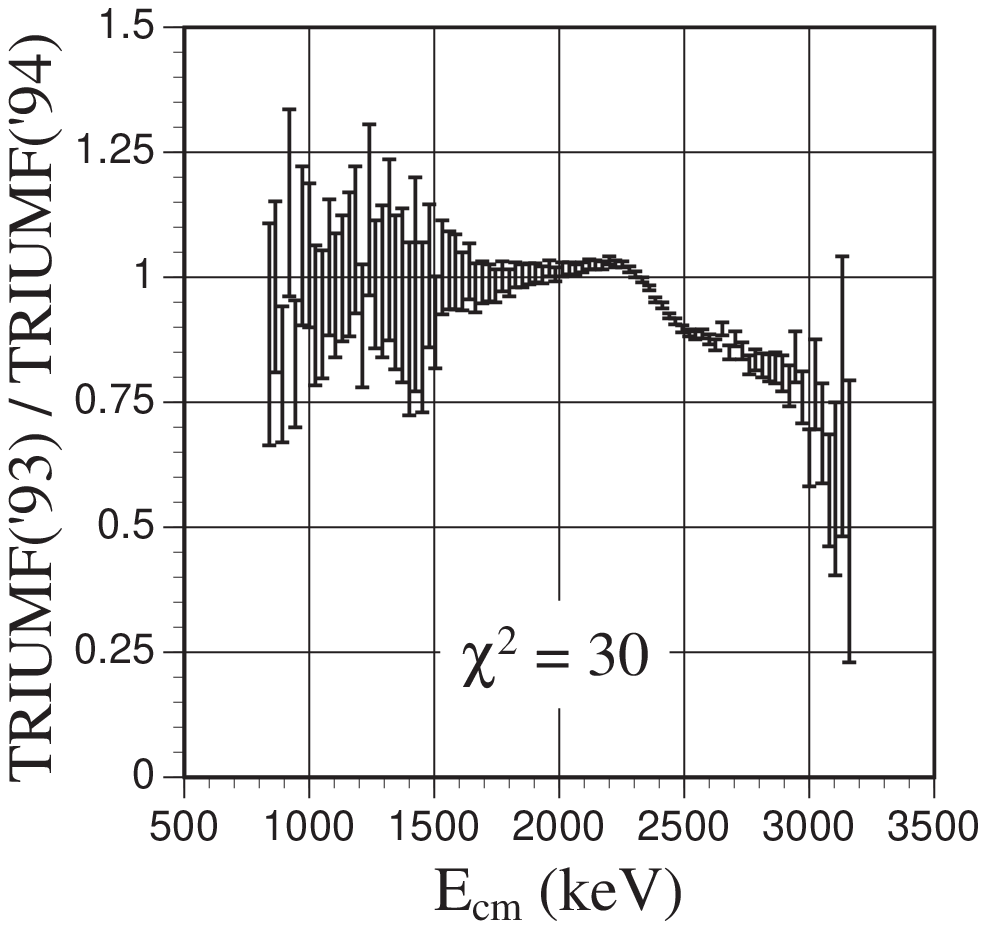,height=3.0in}}

\underline{Fig. 5:}  Ratio of the TRIUMF(94) \cite{Az94} to TRIUMF(93)
\cite{Bu93} data.  The narrower line shape of the TRIUMF(93) data is
understood to be an artifact of the analysis.

\subsection{Conclusions}

We have reviewed the status of both data and analyses pertaining to 
the p-wave astrophysical S-factor of the \c12ag reaction. We observe 
that more recent global R-matrix fit of 
the data on \n16, elastic scattering and \c12ag reaction data
does not allow us to rule out a small S-factor solution and does 
not confirm the strong statement of the TRIUMF collaboration that 
the S-factor is now known.  The sign of the 
interference of the two $1^-$ states in \c12ag data is not 
directly determined by data on the beta-decay of \n16,
and thus this problem remains unsolved and needs to be studied via
additional low energy data on the \c12ag 
reaction itself.

We have improved our original 
data on the beta-delayed alpha-particle
emission of \n16.  A comparison of all four high 
statistics data on \n16 reveals three data sets: the Mainz('71), 
Seattle('95), and the current Yale-UConn('96) that agree with each 
other but disagree with the TRIUMF('94) data (by up to 
a factor of 3). The current situation
with discrepant data on \n16, let alone disagreement on data
on \c12ag capture reaction, and disagreement 
in the extracted S-factor, do
not allow us to conclude that the p-wave S-factor for the \c12ag
reaction is known with an accuracy sufficient for stellar 
evolution models, and we do not confirm neither the TRIUMF data nor
the large S-factor quoted by TRIUMF with a relatively high accuracy.

\section{Acknowledgments}

I would like to acknowledge the work of my graduate student at
Yale University, Dr. Ralph H. France III, on the \n16 data, and
the work of T. Kikuchi, 
Dr. N. Iwasa and Professor
Tohru Motobayashi that performed the data analysis of the \b8
experiments at Rikkyo University.  I would also like 
to acknowledge discussions
and encouragements from Professor Carlos Bertulani,
Professor Gerhard Baur, and my collaborator Professor Thierry Delbar.

\end{document}